\documentclass{PoS}
\usepackage{url}

\title{KLEVER: An experiment to measure BR($K_L\to\pi^0\nu\bar{\nu}$)
  at the CERN SPS}

\ShortTitle{KLEVER: An experiment to measure BR($K_L\to\pi^0\nu\bar{\nu}$)
  at the CERN SPS}

\author{\speaker{Matthew Moulson} for the KLEVER Project\\
        INFN Laboratori Nazionali di Frascati\\
        E-mail: \email{moulson@lnf.infn.it}}

\abstract{
Precise measurements of the branching ratios for the flavor-changing
neutral current decays $K\to\pi\nu\bar{\nu}$ can provide unique constraints
on CKM unitarity and, potentially, evidence for new physics. It is important
to measure both decay modes, $K^+\to\pi^+\nu\bar{\nu}$ and
$K_L\to\pi^0\nu\bar{\nu}$, since different new physics models affect the rates
for each channel differently.
The goal of the NA62 experiment at the CERN SPS
is to measure the BR for the charged channel to within 10\%.
For the neutral channel, the BR has never been measured.
KOTO, an experiment at J-PARC, should have enough data for the
first observation of the decay by about 2021.
We are designing the KLEVER experiment to measure
BR($K_L\to\pi^0\nu\bar{\nu}$) to $\sim$20\% using a high-energy neutral beam
at the CERN SPS starting in Run 4.
The boost from the high-energy beam facilitates the rejection of background
channels such as $K_L\to\pi^0\pi^0$ by detection of the additional photons
in the final state.
On the other hand, the layout poses particular challenges for the
design of the small-angle vetoes, which must reject photons from $K_L$
decays escaping through the beam pipe amidst an intense background from
soft photons and neutrons in the beam. Background from $\Lambda \to n\pi^0$
decays in the beam must also be kept under control.
We present findings from our design studies,
with an emphasis on the challenges faced and the
potential sensitivity for the measurement of
BR($K_L\to\pi^0\nu\bar{\nu}$).}

\FullConference{The 39th International Conference on High Energy Physics (ICHEP2018)\\
		4-11 July, 2018\\
		Seoul, Korea}

\begin{document}

The branching ratios for the decays $K\to\pi\nu\bar{\nu}$
are among the observables in the quark-flavor sector most sensitive to
new physics. The Standard Model (SM) rates for these decays are
${\rm BR}(K^+\to\pi^+\nu\bar{\nu}) = (8.4\pm1.0)\times10^{-11}$ and 
${\rm BR}(K_L\to\pi^0\nu\bar{\nu}) = (3.4\pm0.6)\times10^{-11}$, with 
non-parametric theoretical uncertainties
of about 3.5\% and 1.5\%, respectively~\cite{Buras:2015qea}.
Because these decays are strongly suppressed and their rates are
calculated very precisely in the SM, their BRs are
potentially sensitive to
mass scales of hundreds of TeV, surpassing the sensitivity
of $B$ decays in most SM extensions~\cite{Buras:2014zga}.
Observations of lepton-flavor-universality-violating phenomena are mounting
in the $B$ sector~\cite{Bifani:2018zmi}.
Measurements of the $K\to\pi\nu\bar{\nu}$ BRs are critical
to interpreting the data from rare $B$ decays, and may demonstrate that
these effects are a manifestation of new degrees of freedom such as vector
leptoquarks~\cite{Buttazzo-Bordone}.

The goal of the NA62 experiment at the CERN SPS
is to measure ${\rm BR}(K^+\to\pi^+\nu\bar{\nu})$ to within
10\%~\cite{NA62:2017rwk}.
NA62 has recently presented the preliminary
result ${\rm BR}(K^+\to\pi^+\nu\bar{\nu}) < 14\times10^{-10}$ (95\%CL),
with 1 observed candidate event \cite{Massri:2018xxx}.
This result is based on 2\% of the combined data from running in
2016--2018. Additional running is contemplated for 2021--2022. 
In order to distinguish between different new-physics scenarios,
it is necessary to measure ${\rm BR}(K_L\to\pi^0\nu\bar{\nu})$ as
well~\cite{Buras:2015yca}.
KOTO, an experiment at J-PARC, has
recently obtained the limit
${\rm BR}(K_L\to\pi^0\nu\bar{\nu}) < 3.0\times10^{-9}$ (90\%CL),
with an expected background of $0.42\pm0.18$ events and no candidate
events observed~\cite{Shiomi:2018xxx}.
KOTO should be able to make the
first observation of the $K_L\to\pi^0\nu\bar{\nu}$ decay by the early
2020s~\cite{KOTO:2018xxx}, but a next-generation experiment is
needed in order to measure the BR.
We are designing the KLEVER experiment to use a high-energy neutral beam
at the CERN SPS to achieve 60-event sensitivity for the decay
$K_L\to\pi^0\nu\bar{\nu}$ at the SM BR with an $S/B$ ratio of 1.
\begin{figure}
  \centering
  \includegraphics[width=0.75\textwidth]{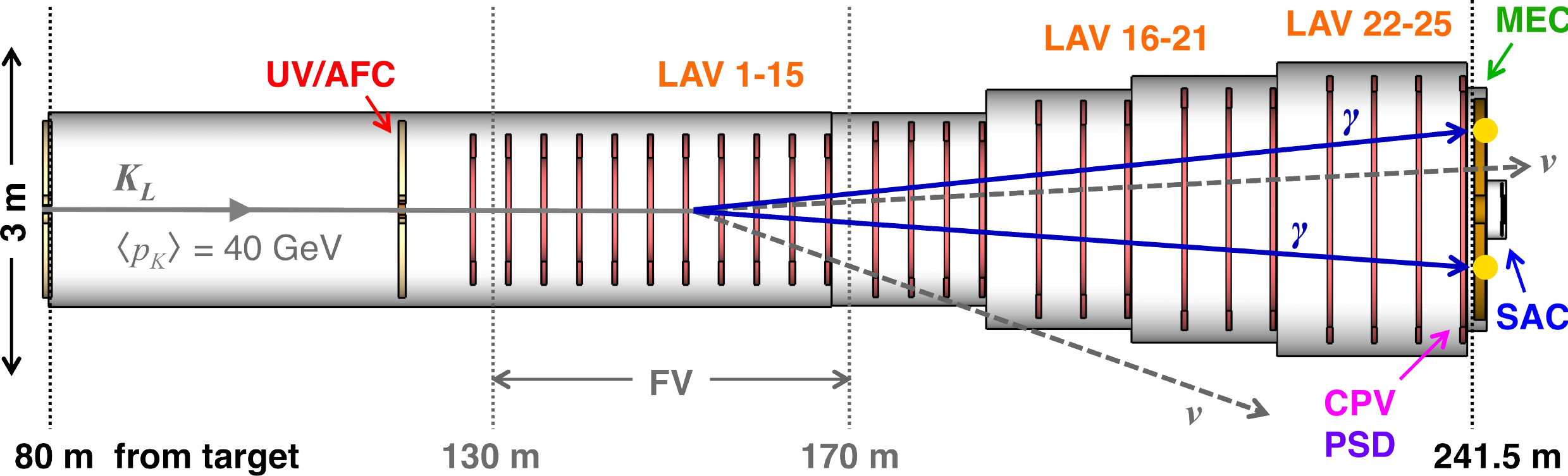}
  \caption{KLEVER experimental apparatus: upstream veto (UV),
    active final collimator (AFC), large-angle photon vetoes (LAV), main
    electromagnetic calorimeter (MEC), small-angle calorimeter (SAC),
    charged particle veto (CPV), pre-shower detector (PSD).}
  \label{fig:exp}
\end{figure}
Data taking would start in LHC Run 4 (2026).
The layout is sketched in Figure~\ref{fig:exp}. 

KLEVER will make use of the 400-GeV SPS proton beam to produce a neutral
secondary beam at an angle of 8 mrad with
a mean $K_L$ momentum of 40 GeV and a $K_L$ yield of
$2\times10^{-5}$ $K_L$ per proton on target (pot).
Collection of 60 SM events in 5 years would require a total
primary flux of $5\times10^{19}$ pot, corresponding to an intensity of
$2\times10^{13}$ protons per pulse (ppp) under NA62-like slow-extraction
conditions.
This is a six-fold increase in the primary intensity relative to NA62, the
feasibility of which is under study.
Preliminary indications are positive:
there is general progress on issues related to the slow extraction
of the needed intensity to the North Area (including duty cycle optimization);
a workable solution for transport of the beam from the primary target to the
experimental beamline has been identified; and the
ventilation in the target and secondary beam cavern appears to be sufficiently
hermetic, obviating the need for potentially expensive upgrades.
The target and dump collimator (TAX) may have to be upgraded or rebuilt.
A four-collimator neutral beamline layout for ECN3 has been developed and
simulation studies with FLUKA and Geant4 are in progress to quantify the
extent and composition of beam halo, muon backgrounds, and sweeping
requirements.
According to the simulation, there are 140 MHz of $K_L$ in the beam
and 440 MHz of neutrons~\cite{vanDijk:2018xxx}.
A tungsten converter in the TAX
followed by sweeping magnets in the beamline eliminates all but 40~MHz
of the photons in the beam with energy greater
than 5 GeV.

Most of the subdetector systems for KLEVER will have to be newly constructed.
Early studies indicated that the NA48 liquid-krypton calorimeter
(LKr) \cite{Fanti:2007vi}
currently used in NA62 could be reused as the MEC. Indeed, the efficiency
and energy resolution of the
LKr appear to be satisfactory for KLEVER. However, the LKr would
measure the event time in KLEVER with 500-ps resolution, while the
total rate of accidental vetoes (dominated by the rate in the SAC)
could be 100 MHz.
We are investigating the possibility of replacing the
LKr with a shashlyk-based MEC patterned on the PANDA FS calorimeter (in turn,
based on the KOPIO calorimeter~\cite{Atoian:2007up}).
We envisage a shashlyk design
incorporating ``spy tiles'' for longitudinal sampling of the shower
development, resulting in additional information for $\gamma/n$ separation.
A first test of this concept was carried out with a prototype detector at
Protvino in April 2018.

The upstream veto (UV), which rejects $K_L\to\pi^0\pi^0$ decays upstream of
the fiducial volume, would use the same shashlyk technology as the MEC.
The active final collimator (AFC), inserted into the hole in the UV for
passage of the beam, is a LYSO collar counter with angled inner surfaces. This
provides the last stage of beam collimation while vetoing photons from $K_L$
that decay in transit through the collimator itself.

Because of the boost from the high-energy beam, it is sufficient for the
large-angle photon vetoes (LAVs) to cover polar angles out to 100 mrad.
The 25 new LAV detectors for KLEVER are
lead/scintillating-tile sampling calorimeters with wavelength-shifting fiber
readout~\cite{Ramberg:2004en}.
Extensive experience with this
type of detector (including in prototype tests for NA62) demonstrates that
the low-energy photon detection efficiency will be sufficient for
KLEVER~\cite{Atiya-Comfort-Ambrosino}.

The small-angle calorimeter (SAC) sits directly in the neutral beam and must
reject photons from $K_L$ decays that would otherwise escape via the
downstream beam exit. The veto efficiency required is not intrinsically
daunting (inefficiency $<1$\% for $5~{\rm GeV} < E_\gamma < 30~{\rm GeV}$
and $<10^{-4}$ for $E_\gamma < 30~{\rm GeV}$; the SAC can be blind for
$E_\gamma < 5~{\rm GeV}$),
but must be attained while maintaining insensitivity to more than 500~MHz
of neutral hadrons in the beam. In addition, the SAC must have good
longitudinal and transverse segmentation to provide $\gamma/n$
discrimination.
An intriguing solution is to construct the SAC as a compact, Si-W sampling
calorimeter with crystalline tungsten tiles as
the absorber material, since coherent interactions of high-energy photons
with a crystal lattice can lead to increased rates of pair conversion
relative to those obtained with amorphous
materials~\cite{Bak-Kimball-Baryshevsky}.
The effect is dependent on photon
energy and incident angle; in the case of KLEVER, one might hope to decrease
the ratio $X_0/\lambda_{\rm int}$ by a factor of 2--3. The same effect could
be used to efficiently convert high-energy photons in the neutral beam to
$e^+e^-$ pairs at the dump collimator for subsequent sweeping, thereby
allowing the use of a thin converter to minimize the scattering of
hadrons from the beam. Both concepts were tested in summer 2018 in the
SPS H2 beam line, in a joint effort together with the AXIAL collaboration.

For the rejection of charged particles, $K_{e3}$ is a benchmark channel
because of its large BR and because the final state electron can be mistaken
for a photon. Simulations indicate that the needed rejection can be achieved
with two staggered planes of charged-particle veto (CPV) each providing 99.5\%
detection efficiency, supplemented by the $\mu^\pm$ and $\pi^\pm$ recognition
capabilities of the MEC (assumed in this case to be equal to those of the LKr)
and the current NA62 hadronic calorimeters and muon vetoes.

Finally, a pre-shower detector (PSD) featuring a 0.5$X_0$ converter and
two planes
of tracking with $\sigma_{x,y} \sim 100~\mu{\rm m}$ (assumed to be large-area
MPGDs) would allow angular reconstruction of at least one $\gamma$ from
$K_L\to\pi^0\pi^0$ events with two lost $\gamma$s in
50\% of cases, providing $\gamma\gamma$ vertex reconstruction
(with the nominal beamline) with $\sigma_z \sim 10$~m or better.
Information from the PSD will be used for bifurcation studies of the
background and for the selection of control samples.  

Simulations of the experiment carried out with fast-simulation techniques
(idealized geometry, parameterized detector response, etc.) show that the
target sensitivity is achievable (60 SM events with $S/B = 1$).
\begin{figure}
  \centering
  \includegraphics[width=0.65\textwidth]{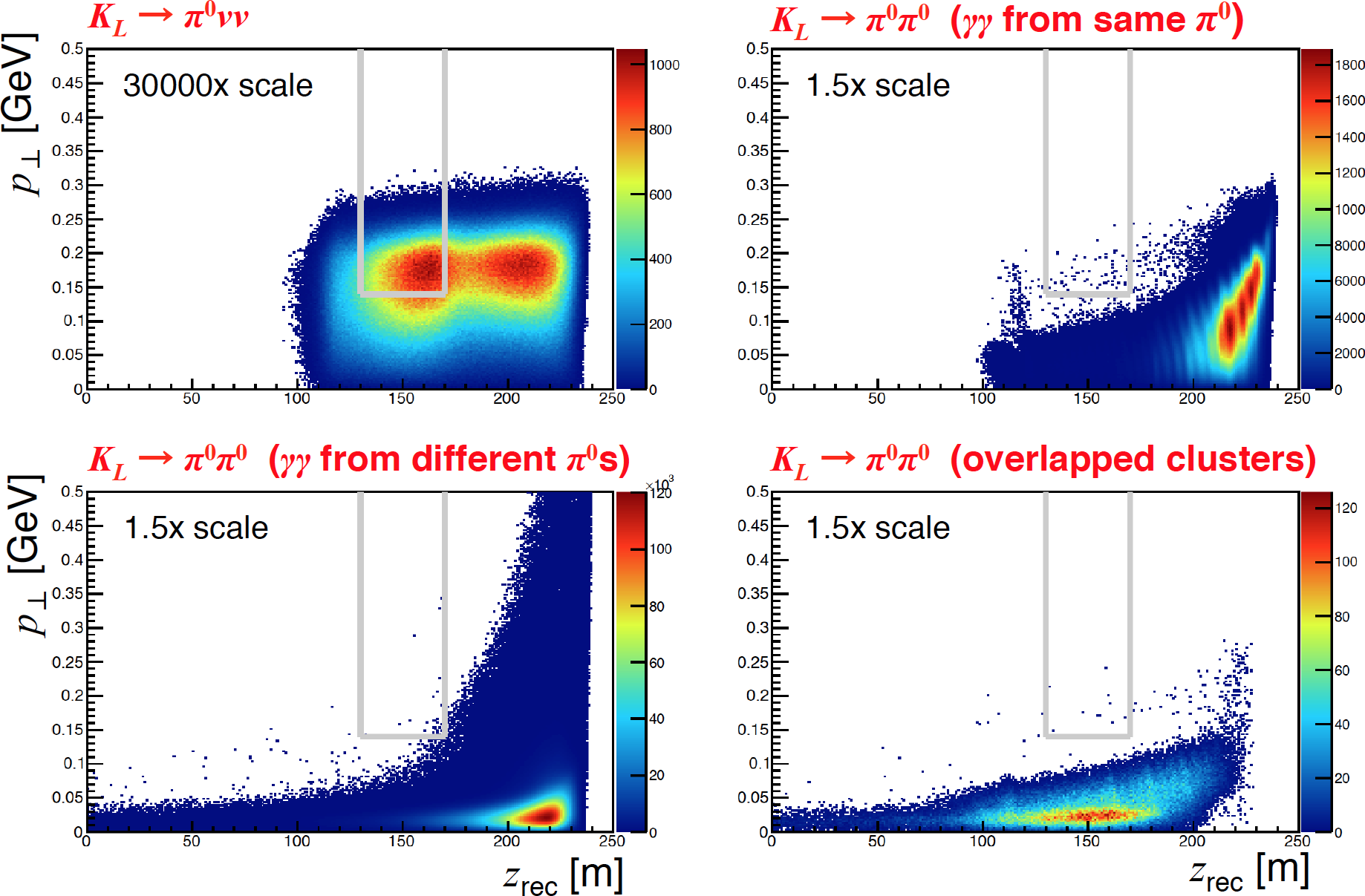}
  \caption{Distributions of events in plane of $(z_{\rm rec}, p_\perp)$
    after basic event selection cuts, from fast MC simulation, for 
    $K_L\to\pi^0\nu\bar{\nu}$ events (top left) and for
    $K_L\to\pi^0\pi^0$ events with two photons from the same
    $\pi^0$ (top right),
    two photons from different $\pi^0$s (bottom left), and
    with two or more indistinguishable overlapping photon clusters
    (bottom right).}
  \label{fig:sel}
\end{figure}
Fig.~\ref{fig:sel} illustrates the scheme for differentiating signal
events from $K_L\to\pi^0\pi^0$ background. Events with exactly two photons
on the MEC and no other activity in the detector are selected. The clusters
on the MEC from both photons must also be more than 35~cm from the beam axis
(this helps to increase the rejection for events with overlapping clusters).
If one or both photons convert in the PSD, the reconstructed vertex
position must be inside the fiducial volume. The plots show the
distributions of the events satisfying these minimal criteria in the
plane of $p_\perp(\gamma\gamma)$ vs.\ $z_{\rm rec}(\gamma\gamma)$
for the $\pi^0$, where the
distance from the $\pi^0$ to the MEC is reconstructed from the transverse
separation of the two photon clusters, assuming that they come from a
$\pi^0$ ($M_{\gamma\gamma} = m_{\pi^0}$). This scheme is far from final and
there is room for improvement with a multivariate analysis, but is does
demonstrate that it should be possible to obtain $S/B\sim1$ with respect
to other $K_L$ decays.
Background
from $\Lambda\to n\pi^0$ and from decays with charged particles is assumed
to be eliminated on the basis of studies with more limited statistics.
Besides the length from the target to the fiducial volume and the choice
of production angle carefully optimized to balance $K_L$ flux against
the need to soften the $\Lambda$ momentum spectrum, background from
$\Lambda\to n\pi^0$ ($p^* = 104$~MeV) can be effectively eliminated
by cuts on $p_\perp$ and in the $\theta$ vs. $p$ plane for the $\pi^0$.  

An effort is underway to develop a comprehensive simulation
and use it to validate the results obtained so far. Of particular note,
backgrounds from radiative $K_L$ decays, cascading hyperon decays, and
beam-gas interactions remain to be studied, and the neutral-beam halo
from our more detailed FLUKA simulations needs to be incorporated into
the simulation of the experiment.
Preliminary studies indicate that
the hit and event rates on most of the detectors are on the order of
a few tens of MHz, a few times larger than in NA62, with the notable
exception of the SAC, which will require an innovative readout solution
to handle rates of 100 MHz.

KLEVER would aim to start data taking in LHC Run 4.
To be ready for the 2026 start date,
detector construction would have to begin by 2021 and be ready for
installation by 2025, leaving three years from the present for design
consolidation and R\&D. Many institutes currently participating in NA62
have expressed support for and interest in the KLEVER project.
Input to the update process for the European Strategy for Particle Physics
and an Expression of Interest to the SPS program committee (SPSC) are in
preparation. Successfully carrying out the KLEVER experimental program
will require the involvement of new institutions and groups, and we are
actively seeking to expand the collaboration at present.

\end{document}